\documentstyle[tighten,aps]{revtex}
\input{psfig.sty}

\newcommand{\la}{\left\langle}
\newcommand{\ra}{\right\rangle}
\newcommand{\EPL}{{\it Europhys.~Lett.~}}
\newcommand{\PRL}{{\it Phys.~Rev.~Lett.~}}
\newcommand{\PR}{{\it Phys.~Rev.~}}
\newcommand{\JCP}{{\it J.~Chem.~Phys.~}}
\newcommand{\JPCM}{{\it J.~Phys.: Condens.~Matter~}}
\newcommand{\MP}{{\it Mol.~Phys.~}}
\newcommand{\JCIS}{{\it J.~Coll.~Int.~Sci.~}}

\begin{document}
\draft

\author{A. R. Denton$^*$}
\address
{Department of Physics, Acadia University, Wolfville, NS, Canada B0P 1X0}

\title{Effective Interactions and Volume Energies in 
Charged Colloids: \\ Linear Response Theory}

\date{\today}
\maketitle

\begin{abstract}
Interparticle interactions in charge-stabilized colloidal suspensions, 
of arbitrary salt concentration, are described at the level of 
effective interactions in an equivalent one-component system.
Integrating out from the partition function the degrees of freedom of 
all microions, and assuming linear response to the macroion charges, 
general expressions are obtained for both an effective electrostatic 
pair interaction and an associated microion volume energy. 
For macroions with hard-sphere cores, the effective interaction is of 
the DLVO screened-Coulomb form, but with a modified screening constant 
that incorporates excluded volume effects. 
The volume energy -- a natural consequence of the one-component reduction 
-- contributes to the total free energy and can significantly influence 
thermodynamic properties in the limit of low-salt concentration.
As illustrations, the osmotic pressure and bulk modulus are computed and 
compared with recent experimental measurements for deionized suspensions. 
For macroions of sufficient charge and concentration, it is shown that 
the counterions can act to soften or destabilize colloidal crystals.
\end{abstract}

\bigskip
\bigskip




\section{Introduction}

More than a century ago, it was recognized that most colloidal particles 
carry an electric charge~\cite{Hunter}.  Colloidal macroions -- typically 
$1-1000$ nm in diameter -- may acquire charges from surface dissociation of 
counterions, adsorption of salt ions from solution, or creation of defects 
in crystal lattices.  Electrostatic repulsion between macroions suspended in 
a molecular fluid is one of the two chief mechanisms by which colloidal 
suspensions may be stabilized against coagulation induced by attractive 
van der Waals forces.

Charge-stabilized colloidal suspensions exist in a wide variety of forms.
Familiar examples include clay minerals (relevant to mineralogy, agriculture, 
and the paper industry), paints, inks, and solutions of charged micelles. 
Further examples are synthetic latex or silica microspheres~\cite{Pusey}, 
which may self-assemble, if sufficiently monodisperse, into ordered crystals.
Aside from providing valuable model systems for fundamental studies of 
condensed matter, colloidal crystals exhibit unique optical properties 
that have inspired a number of recent applications, {\it e.g.}, nanosecond 
optical switches~\cite{Asher1}, chemical sensors~\cite{Asher2}, and 
photonic band gap materials~\cite{Vos}.  

Despite the considerable and growing technological importance of charged 
colloids, progress in predicting macroscopic properties is limited by 
an incomplete understanding of interparticle interactions.
Most theoretical treatments of electrostatic interactions are rooted in 
the landmark theory of Derjaguin, Landau, Verwey, and Overbeek 
(DLVO)~\cite{DLVO}.  The DLVO theory describes the bare Coulomb interactions 
between macroions as screened by the surrounding microions 
(counterions and salt ions).  The resulting 
screened-Coulomb pair potential accounts -- at least qualitatively -- for 
a range of observed phenomena, including the dependence of coagulation rate 
on counterion valence and trends in phase stability with varying salt 
concentration.  Recently, interest in colloidal interactions has intensified 
as a result of accumulating experimental evidence for apparent 
long-range attractions between macroions~\cite{Grier,Ise}. 

A rigorous statistical mechanical treatment of the multi-component mixture 
of macroions, counterions, salt ions, and solvent molecules is a daunting
task.  Interactions in such complex systems are therefore usually treated 
at the level of {\it effective} interactions.
Tracing out from the partition function statistical degrees of freedom 
associated with all but a single component, the mixture is formally mapped 
onto an equivalent one-component system of ``pseudo-particles" governed by
an effective state-dependent interaction~\cite{Rowlinson}.
Effective interactions in charge-stabilized colloids have been modeled
by a variety of techniques, including Poisson-Boltzmann cell 
models~\cite{Alexander,Hansen,Levin} density-functional 
theory~\cite{Lowen,vRH,vRDH,vRE,Graf,Goulding}, Monte Carlo and 
Molecular Dynamics simulation~\cite{Stevens,Damico,Allahyarov1,Allahyarov2}, 
and powerful {\it ab initio} methods~\cite{Lowen,abinitio}.

Here we adopt an alternative approach, recently proposed by Silbert 
{\it et al.}~\cite{Silbert1,Silbert2}, which exploits analogies between 
charged colloids and metals.  
Performing a classical trace over microion degrees of freedom and
treating the electrostatic response of the microions to the macroions 
within second-order perturbation theory, leads to an effective pair 
interaction between pseudo-macroions and an associated volume energy.
The volume energy, which contributes to the total free energy, must be 
included when calculating thermodynamic properties of charged colloids 
modeled by an effective pair potential~\cite{vRH,vRDH,vRE,Graf,Silbert1,DL}.  
Noting the correspondences, microion $\leftrightarrow$ electron and 
macroion $\leftrightarrow$ metallic ion, the response approach is 
the colloidal equivalent of the widely-used pseudo-potential theory of 
metals~\cite{PPT,Hafner,AS}.

In previous work~\cite{Denton1}, the response approach was extended to
finite-sized macroions in deionized suspensions.
This paper generalizes the theory to the case of arbitrary salt concentration, 
consistently taking into account (1) the volume excluded to the microions 
by the macroion hard cores and (2) the response of both counterions and 
salt ions to the macroion charges.
The next section begins with a brief review of the response theory and 
then outlines our extensions of the theory.  Section \ref{results} presents 
the main results -- obtained within a linear response approximation -- for 
an effective pair potential acting between pseudo-macroions and an associated 
volume energy, both of which consistently incorporate excluded volume effects.  
The influence of the volume energy on thermodynamic properties is illustrated 
by calculations of the osmotic pressure and bulk modulus.  Comparisons
with experimental data show that the counterions contribute a substantial
fraction of the osmotic pressure and can soften or destabilize colloidal
crystals.  Finally, in Sec.~\ref{conclusions} we summarize and conclude.

\section{Theory}

\subsection{The Model}

Within the ``primitive" model, the system comprises $N_{\rm m}$ charged 
hard-sphere macroions of diameter $\sigma$ and charge $-Ze$ ($e$ being 
the elementary charge) and $N_{\rm c}$ point counterions of charge $ze$ 
suspended in an electrolyte solvent.  Global charge neutrality constrains 
macroion and counterion numbers according to $ZN_{\rm m}= zN_{\rm c}$.
Each macroion is assumed to carry a fixed charge, uniformly distributed 
over its surface.  
The solvent hosts $N_{\rm s}$ pairs of salt ions in a uniform dielectric 
fluid characterized entirely by a dielectric constant $\epsilon$. 
For notational simplicity, we assume a symmetric 1:1 electrolyte, 
consisting of $N_{\rm s}$ point ions of charge $ze$ and 
$N_{\rm s}$ of charge $-ze$ ({\it i.e.}, same valence as counterions).
The microions thus number $N_{\rm +}=N_{\rm c}+N_{\rm s}$ positive 
and $N_{\rm -}=N_{\rm s}$ negative, for a total of
$N_{\mu}=N_{\rm c}+2N_{\rm s}$ . 
The system occupies a total volume $V$ at temperature $T$ and fixed salt 
concentration maintained by exchange of salt ions, through 
a semi-permeable membrane, with a salt reservoir.

Denoting macroion and microion coordinates by $\{{\bf R}\}$ and 
$\{{\bf r}\}$, respectively, the Hamiltonian of the system may be expressed 
in the general form
\begin{equation}
H(\{{\bf R}\},\{{\bf r}\})~=~H_{\rm m} + H_{\mu} + 
H_{\rm m+}+ H_{\rm m-}.
\label{H}
\end{equation}
The first two terms on the right side of Eq.~(\ref{H}) denote Hamiltonians 
for macroions and microions, respectively. 
Assuming the only relevant interactions to be steric and electrostatic, 
the bare macroion Hamiltonian, $H_{\rm m}$, is given by
\begin{equation}
H_{\rm m}~=~K_{\rm m} + {1\over{2}}\sum_{{i,j=1}\atop{(i\neq j)}}^{N_{\rm m}}
\left[v_{\rm HS}(|{\bf R}_i-{\bf R}_j|)+v_{\rm mm}(|{\bf R}_i-{\bf R}_j|)
\right],
\label{Hm}
\end{equation}
$K_{\rm m}$ being the kinetic energy of the macroions, 
$v_{\rm HS}(|{\bf R}_i-{\bf R}_j|)$ the hard-sphere pair interaction 
between macroion cores, and $v_{\rm mm}(r)=Z^2e^2/\epsilon r$ the 
bare Coulomb interaction between a pair of macroions separated by 
center-to-center distance $r>\sigma$.
Similarly, the microion Hamiltonian takes the form
\begin{eqnarray}
H_{\mu}~&=&~K_{\mu}~+~\sum_{i=1}^{N_{\mu}}~\sum_{j=1}^{N_{\rm m}} 
v_{\rm HS}(|{\bf r}_i-{\bf R}_j|)~+~{1\over{2}}\sum_{{i,j=1}\atop{(i\neq j)}}^
{N_{\rm +}} v_{\rm ++}(|{\bf r}_i-{\bf r}_j|) \nonumber \\
 &+& \sum_{i=1}^{N_{\rm +}}~\sum_{j=1}^{N_{\rm -}} 
v_{\rm +-}(|{\bf r}_i-{\bf r}_j|)~+~
{1\over{2}}\sum_{{i,j=1}\atop{(i\neq j)}}^{N_{\rm -}} 
v_{\rm --}(|{\bf r}_i-{\bf r}_j|), 
\label{Hmu}
\end{eqnarray}
where $K_{\mu}$ is the microion kinetic energy, 
$v_{\rm HS}(|{\bf r}_i-{\bf R}_j|)$ is the hard-sphere interaction 
between a point microion and a macroion core, and
$v_{\rm ++}(r)=v_{\rm --}(r)=-v_{\rm +-}(r)=z^2e^2/\epsilon r$ 
is the microion-microion Coulomb interaction.
The last two terms in Eq.~(\ref{H}) are the macroion-microion electrostatic 
interaction energies, given by
\begin{equation}
H_{\rm m\pm}~=~\sum_{i=1}^{N_{\rm \pm}}\sum_{j=1}^{N_{\rm m}} 
v_{\rm m\pm}(|{\bf r}_i-{\bf R}_j|),
\label{Hm+-1}
\end{equation}
where $v_{\rm m\pm}(r)$ denotes the macroion-microion electrostatic 
pair interaction.  For later reference, we note that Eq.~(\ref{Hm+-1})
also may be expressed in the form
\begin{equation}
H_{\rm m\pm}~=~\int{\rm d}{\bf r} \int{\rm d}{\bf R} \rho_{\rm \pm}({\bf r}) 
\rho_{\rm m}({\bf R}) v_{\rm m\pm}(|{\bf r}-{\bf R}|),
\label{Hm+-2}
\end{equation}
where 
\begin{equation}
\rho_{\rm \pm}({\bf r}) \equiv \sum_{i=1}^{N_{\rm \pm}}
\delta({\bf r}-{\bf r}_i), \qquad\qquad
\rho_{\rm m}({\bf R}) \equiv \sum_{j=1}^{N_{\rm m}} 
\delta({\bf R}-{\bf R}_j)
\label{rho}
\end{equation}
are the microion and macroion density operators, whose Fourier transforms are 
\begin{equation}
\hat\rho_{\rm \pm}({\bf k})=
\sum_{i=1}^{N_{\rm\pm}}\exp(i{\bf k}\cdot{\bf r}_i), \qquad\qquad
\hat\rho_{\rm m}({\bf k})=\sum_{j=1}^{N_{\rm m}}\exp(i{\bf k}\cdot{\bf R}_j).
\label{rhok}
\end{equation}
Although $v_{\rm m\pm}(r)$ has the Coulomb form
outside the macroion core radius, inside the core it has no unique 
definition.  Thus, following van Roij and Hansen~\cite{vRH}, we are free 
to choose $v_{\rm m\pm}(r)$ to be a constant for $r<\sigma/2$ and take
\begin{equation}
v_{\rm m\pm}(r)~=~\left\{ \begin{array}
{l@{\quad\quad}l}
\frac{\displaystyle \mp Zze^2}{\displaystyle \epsilon r}, & r>\sigma/2 \\
\frac{\displaystyle \mp Zze^2}{\displaystyle \epsilon\sigma/2}\alpha, 
& r<\sigma/2, \end{array} \right.
\label{vm+-r}
\end{equation}
where the parameter $\alpha$ will be specified (Sec.~III.C) to ensure that 
the microion densities vanish within the core.

\subsection{Reduction to an Equivalent One-Component System}

With the Hamiltonian specified, we now turn to a statistical mechanical 
description of the system, our ultimate goal being the free energy.  
The canonical partition function is given by
\begin{equation}
{\cal Z}~=~\la\la\exp(-H/k_{\rm B}T)\ra_{\mu}\ra_{\rm m},
\label{part1}
\end{equation}
the angular brackets symbolizing classical traces over microion and
macroion degrees of freedom.
Following standard treatments originating from the theory of simple 
metals~\cite{Hafner,AS,HM}, we now reduce the two-component macroion-microion 
mixture to an equivalent one-component system by performing a restricted
trace over microion coordinates, keeping macroion coordinates fixed.  
Thus, without approximation in this purely classical system,
\begin{equation}
{\cal Z}~=~\la\exp(-H_{\rm eff}/k_{\rm B}T)\ra_{\rm m},
\label{part2}
\end{equation}
where $H_{\rm eff}\equiv H_{\rm m}+F_{\mu}$ is the effective Hamiltonian of 
a one-component system of pseudo-macroions, and where
\begin{equation}
F_{\mu}~\equiv~-k_{\rm B}T\ln\la 
\exp\left[-(H_{\mu}+H_{\rm m+}+H_{\rm m-})/k_{\rm B}T\right]\ra_{\mu}
\label{Fmu1}
\end{equation}
may be physically interpreted as the free energy of a nonuniform gas of 
microions in the midst of macroions fixed at positions ${\bf R}_i$.
Formally adding to and substracting from $H$ the energy, $E_{\rm b}$, 
of a uniform background having a charge equal to that of the macroions,  
Eq.~(\ref{Fmu1}) may be recast in the form
\begin{equation}
F_{\mu}~=~-k_{\rm B}T\ln\la\exp\left[-(H_{\mu}' +
H_{\rm m+}'+H_{\rm m-}')/k_{\rm B}T\right]\ra_{\mu},
\label{Fmu2}
\end{equation}
where $H_{\mu}'=H_{\mu}+E_{\rm b}$ and
$H_{\rm m\pm}'=H_{\rm m\pm}-E_{\rm b}/2$.  
The advantage of this simple manipulation is that $H_{\mu}'$ is
the Hamiltonian of a classical, two-component plasma of microions, 
in a uniform compensating background,  
in the presence of {\it neutral} hard-sphere macroions. 
In order that the plasma be free of infinities associated with the long-range 
Coulomb interaction, the background must occupy the same volume as 
the microions.  The background is thus excluded -- along with the microions -- 
from the macroion cores.  The microions and background then jointly occupy 
a {\it free} volume $V'\equiv V(1-\eta)$, which is the total volume reduced by 
the volume fraction of the macroion cores, $\eta=(\pi/6)(N_{\rm m}/V)\sigma^3$.

The background energy is given explicitly by~\cite{HM}
\begin{eqnarray}
E_{\rm b}~&=&~{1\over{2}}(n_{\rm +}-n_{\rm -})^2\int_{V'}
{\rm d}{\bf r} \int_{V'}{\rm d}{\bf r}'{{z^2e^2}\over{\epsilon 
|{\bf r}-{\bf r}'|}}~-~\sum_{i=1}^{N_{\rm m}}\int_{V'}
{\rm d}{\bf r} {{(n_{\rm +}-n_{\rm -})Zze^2}\over{\epsilon 
|{\bf r}-{\bf R}_i|}} \nonumber \\
 ~&=&~-{1\over{2}}(N_{\rm +}-N_{\rm -})(n_{\rm +}-n_{\rm -})\hat v_{\rm ++}(0),
\label{Vb}
\end{eqnarray}
where $n_{\rm \pm}=N_{\rm \pm}/V'=n^{(0)}_{\rm \pm}/(1-\eta)$ are 
the {\it effective} mean densities of microions in the volume not occupied
by the macroion cores and $n^{(0)}_{\rm \pm}=N_{\rm \pm}/V$ are
the {\it nominal} mean densities.
For later reference, we also define $n_{\rm s}=n_{\rm -}$ and 
$n_{\rm c}=n_{\rm +}-n_{\rm -}$ as the effective densities of 
salt-ion pairs and counterions, respectively. 
In Eq.~(\ref{Vb}), $\hat v_{\rm ++}(0)$, defined by
\begin{equation}
\hat v_{\rm ++}(0)~=~\int_{V'}{\rm d}{\bf r}\frac{z^2e^2}{\epsilon r}
~=~\lim_{k\to 0}\left({{4\pi z^2e^2}\over{\epsilon k^2}}\right),
\label{v++0}
\end{equation}
is the $k \to 0$ limit of the Fourier transform of $v_{\rm ++}(r)$.
Although formally infinite, $E_{\rm b}$ will be seen below to be identically 
cancelled by compensating infinities in $H_{\mu}$ and $H_{\rm m\pm}$.  

\subsection{Linear Response Approximation}
 
The theory presented thus far is exact, within the primitive model.  
The challenge remains to calculate the microion free energy [Eq.~(\ref{Fmu1})].
One proposed strategy~\cite{vRH} invokes density-functional theory to 
approximate $F_{\mu}$, regarded as a functional of the microion densities, by
performing a functional Taylor-series expansion about a uniform microion plasma.
An alternative strategy~\cite{Silbert1,Silbert2}, inspired by the
pseudo-potential theory of metals, is to formally regard $H_{\rm m\pm}'$ 
as ``external" potentials acting upon a microion plasma and then approximate 
$F_{\mu}$ by perturbation theory.  
Following the second strategy, we write~\cite{HM}
\begin{equation}
F_{\mu}~=~F_{\rm p}~+~\int_0^1{\rm d}\lambda\left(\la H_{\rm m+}'
\ra_{\lambda}+\la H_{\rm m-}'\ra_{\lambda}\right),
\label{Fmu3}
\end{equation}
where 
\begin{equation}
F_{\rm p}~=~-k_{\rm B}T\ln\la\exp(-H_{\mu}'/k_{\rm B}T)\ra_{\mu}
\label{Fp1}
\end{equation}
is the free energy of the reference microion plasma, occupying a volume 
$V'$, in the presence of neutral hard-sphere macroions.
The integral over $\lambda$ in Eq.~(\ref{Fmu3}) corresponds physically to
an adiabatic charging of the macroions from neutral to fully-charged spheres.
The ensemble average $\la~\ra_{\lambda}$ represents an average 
with respect to the distribution function of a system whose macroions 
carry a charge $\lambda Z$.

Further progress is facilitated by expressing $\la H_{\rm m\pm}'\ra_{\lambda}$ 
in terms of Fourier components of the macroion and microion densities and 
of the macroion-microion interaction.  From Eqs.~(\ref{Hm+-2})-(\ref{rhok}), 
we have
\begin{equation}
\la H_{\rm m\pm}'\ra_{\lambda}~=~
\frac{1}{V'} \sum_{{\bf k}\neq 0} \hat v_{\rm m\pm}(k) 
\la\hat\rho_{\rm \pm}({\bf k})\ra_{\lambda} \hat\rho_{\rm m}(-{\bf k}) + 
\frac{1}{V'} \lim_{k\to 0}\left[ \hat v_{\rm m\pm}(k)
\la\hat\rho_{\rm \pm}({\bf k})\ra_{\lambda} \hat\rho_{\rm m}(-{\bf k})\right] - 
E_{\rm b}/2.
\label{Hm+-3}
\end{equation}
Evidently $\la H_{\rm m\pm}'\ra_{\lambda}$ depends through
$\hat\rho_{\rm \pm}({\bf k})$ upon the
response of the microions to the macroion charge density.
Regarding the macroion charge as imposing an external potential on the 
microions, and assuming that the microion densities respond {\it linearly}
to this potential, the Fourier components of the microion densities 
appearing in Eq.~(\ref{Hm+-3}) may be expressed in the form
\begin{equation}
\la\hat\rho_{\rm +}({\bf k})\ra_{\lambda}~=~\lambda\left[\chi_{\rm ++}(k)
- \chi_{\rm +-}(k)\right] \hat v_{\rm m+}(k)
\hat\rho_{\rm m}({\bf k}),\qquad k\neq 0,
\label{rho+}
\end{equation}
and
\begin{equation}
\la\hat\rho_{\rm -}({\bf k})\ra_{\lambda}~=~\lambda\left[\chi_{\rm +-}(k)
- \chi_{\rm --}(k)\right] \hat v_{\rm m+}(k)
\hat\rho_{\rm m}({\bf k}),\qquad k\neq 0,
\label{rho-}
\end{equation}
where $\chi_{\rm\pm\pm}(k)$ are the linear response functions 
of the reference two-component microion plasma and where we have used 
the symmetry relations $\chi_{\rm +-}(k)=\chi_{\rm -+}(k)$ and
$\hat v_{\rm m+}(k)=-\hat v_{\rm m-}(k)$.
Note that for $k=0$ there is no response, since 
$\hat\rho_{\rm \pm}(0)=N_{\rm \pm}$, as determined by 
the fixed numbers of microions.
Substituting Eqs.~(\ref{rho+}) and (\ref{rho-}) into Eq.~(\ref{Hm+-3}) 
and this in turn into Eq.~(\ref{Fmu3}), and integrating over $\lambda$, 
the microion free energy is given
to second order in the macroion-microion interaction by
\begin{eqnarray}
F_{\mu}~&=&~F_{\rm p}~+~\frac{1}{2V'}\sum_{{\bf k}\neq 0}
\left[\chi_{\rm ++}(k)-2\chi_{\rm +-}(k)+\chi_{\rm --}(k)\right]
\left[\hat v_{\rm m+}(k)\right]^2
\hat\rho_{\rm m}({\bf k})\hat\rho_{\rm m}(-{\bf k}) \nonumber \\
 ~&+&~N_{\rm m}(n_{\rm +}-n_{\rm -})\lim_{k\to 0}\left[\hat v_{\rm m+}(k)\right]
- E_{\rm b},
\label{Fmu4}
\end{eqnarray}
where again we have used the relation $\hat v_{\rm m+}(k)=-\hat v_{\rm m-}(k)$.
Correspondingly, the effective Hamiltonian takes the form
\begin{eqnarray}
H_{\rm eff}&=&K_{\rm m}+{1\over{2}}\sum_{{i,j=1}\atop{i\neq j}}^{N_{\rm m}}
v_{\rm HS}(|{\bf R}_i-{\bf R}_j|)+
\frac{1}{2V'}\sum_{\bf k}\hat v_{\rm mm}(k)
\left[\hat\rho_{\rm m}({\bf k})\hat\rho_{\rm m}(-{\bf k})-N_{\rm m}\right] 
\nonumber \\
&+& F_{\rm p} + \frac{1}{2V'}\sum_{\bf k\neq 0}\chi(k)
\left[\hat v_{\rm m+}(k)\right]^2\hat\rho_{\rm m}({\bf k})\hat\rho_{\rm m}
(-{\bf k}) + N_{\rm m}(n_{\rm +}-n_{\rm -})\lim_{k\to 0}
\left[\hat v_{\rm m+}(k)\right] - E_{\rm b},
\label{Heff1}
\end{eqnarray}
where we have defined 
\begin{equation}
\chi(k)\equiv \chi_{\rm ++}(k)-2\chi_{\rm +-}(k)+\chi_{\rm --}(k).
\label{chi1}
\end{equation}
Now rearranging terms, Eq.~(\ref{Heff1}) may be restructured and written 
in the formally simpler form
\begin{eqnarray}
H_{\rm eff}&=&K_{\rm m}+{1\over{2}}\sum_{{i,j=1}\atop{i\neq j}}^{N_{\rm m}}
v_{\rm HS}(|{\bf R}_i-{\bf R}_j|) +
\frac{1}{2V'}\sum_{\bf k}\hat v_{\rm eff}(k)
\left[\hat\rho_{\rm m}({\bf k})\hat\rho_{\rm m}(-{\bf k})-N_{\rm m}\right] 
+ E_{\rm 0} \nonumber\\
&=&K_{\rm m} + {1\over{2}}\sum_{{i,j=1}\atop{i\neq j}}^{N_{\rm m}}
\left[v_{\rm HS}(|{\bf R}_i-{\bf R}_j|) + 
v_{\rm eff}(|{\bf R}_i-{\bf R}_j|)\right] + E_{\rm 0},
\label{Heff2}
\end{eqnarray}
where
\begin{equation}
v_{\rm eff}(r)~=~v_{\rm mm}(r)~+~v_{\rm ind}(r)
\label{veffr1}
\end{equation}
has the physical interpretation of an {\it effective} electrostatic 
pair potential between pseudo-macroions, which is the sum of 
the bare Coulomb potential and an {\it induced} potential whose
Fourier transform is
\begin{equation}
\hat v_{\rm ind}(k)~=~\chi(k)\left[\hat v_{\rm m+}(k)\right]^2.
\label{vindk1}
\end{equation}
The final term in Eq.~(\ref{Heff2}) is the {\it volume energy}, 
\begin{equation}
E_{\rm 0}~=~F_{\rm p} + {{N_{\rm m}}\over{2}}\lim_{r\to 0} v_{\rm ind}(r)
+ N_{\rm m}(n_{\rm +}-n_{\rm -})\lim_{k\to 0}\left[-{{z}\over{2Z}}
\hat v_{\rm ind}(k) + \hat v_{\rm m+}(k)\right] - E_{\rm b},
\label{Eo1}
\end{equation}
which is a natural byproduct of the reduction to 
an equivalent one-component system.  
Although it has no explicit dependence on the macroion coordinates (see below), 
$E_{\rm 0}$ evidently depends on the mean density of macroions and therefore
can contribute significantly to the total free energy of the system.
In passing, we note that the above expressions for the effective pair potential 
and the volume energy are analogous to expressions appearing in
the pseudo-potential theory of metals~\cite{Hafner,AS,HM,Finnis} if one 
substitutes for $F_{\rm p}$ and $\chi(k)$, respectively, the energy and 
linear response function of the homogeneous electron gas 
(in the presence of a compensating background), 
and for $\hat v_{\rm m+}(k)$ the electron-ion pseudo-potential.

Summarizing thus far, we have adopted the primitive model of charged colloids, 
formally reduced the macroion-microion mixture to an equivalent
one-component system of pseudo-macroions, and applied a linear response 
approximation to the microion density, to obtain expressions for 
an effective electrostatic pair interaction [Eqs.~(\ref{veffr1}) and 
(\ref{vindk1})] and an associated volume energy [Eq.~(\ref{Eo1})].
Practical calculations still require explicit specification of 
(1) the reference plasma free energy $F_{\rm p}$, 
(2) the plasma linear response functions $\chi_{\rm\pm\pm}(k)$, and 
(3) the macroion-microion interaction $\hat v_{\rm m+}(k)$.
In the next section we consider each of these in turn.

\section{Results and Discussion}\label{results}

\subsection{Reference Microion Plasma}\label{plasma}

The free energy of the two-component reference plasma may be expressed as
\begin{equation}
F_{\rm p}~=~F_{\rm id}~+~F_{\rm corr}~+~F_{\rm cc}-E_{\rm b},
\label{Fp2}
\end{equation}
where $F_{\rm id}$ and $F_{\rm corr}$ are the ideal-gas and correlation 
contributions and $F_{\rm cc}$ is the energy associated with 
Coulomb pair interactions between microions. 
It is important to emphasize that 
by associating the hard-sphere part of the total macroion-microion interaction 
with the microion Hamiltonian [Eq.~(\ref{Hmu})] -- required, since response 
theory does not apply to hard-sphere interactions -- the reference microion
plasma is implicitly restricted to the free volume outside of 
the macroion cores.  As a consequence, the plasma is not strictly uniform, 
since the boundary conditions, imposed by the macroion surfaces, 
may induce nonuniformity.  In general, the ideal-gas free energy is given by
\begin{equation}
\beta F_{\rm id}~=~\int{\rm d}{\bf r}\rho_{\rm +}^{(0)}({\bf r})
\left(\ln[\rho_{\rm +}^{(0)}({\bf r})\Lambda^3]-1\right)~+~
\int{\rm d}{\bf r}\rho_{\rm -}^{(0)}({\bf r})
\left(\ln[\rho_{\rm -}^{(0)}({\bf r})\Lambda^3]-1\right),
\label{Fid}
\end{equation}
where $\beta\equiv 1/k_{\rm B}T$, $\Lambda$ is the microion thermal de Broglie 
wavelength, and $\rho_{\rm\pm}^{(0)}({\bf r})$ are the nonuniform densities 
of positive and negative microions in an external field due to the macroion 
cores (but {\it not} their electric fields).

Now, for typical macroion charges and concentrations, counterion concentrations 
are in the $\mu$M ($10^{-6}$ mol/l) range.  If the salt concentration also 
falls in this range, then microion concentrations are low enough that 
the plasma is essentially uniform. 
In this case, $F_{\rm cc}\simeq E_{\rm b}$ and the last two terms 
in Eq.~(\ref{Fp2}) cancel each other. 
Furthermore, a plasma of such low concentration is weakly coupled, 
with coupling parameter $\Gamma\equiv z^2e^2/\epsilon k_{\rm B}Ta_{\mu} << 1$, 
where $a_{\mu}=(3/4\pi n_{\mu})^{1/3}$ is the microion sphere radius and
$n_{\mu}=n_{\rm +}+n_{\rm -}$ is the total microion number density.
(In sharp contrast, electron plasmas in metals
are typically characterized by $\Gamma >> 1$.)
The correlation free energy per microion then may be approximated by 
the Abe expansion~\cite{Abe}
\begin{equation}
\frac{\beta F_{\rm corr}}{N_{\mu}}~=~
-\frac{1}{\sqrt{3}}\Gamma^{3/2}~+~O(\Gamma^3),
\label{Abe}
\end{equation}
the leading term being the Debye-H\"uckel approximation~\cite{LL}.
Thus, at low salt concentrations, if nonuniformities and correlations 
are ignored~\cite{vRH,Graf,DL}, a reasonable approximation for
the free energy of the microion plasma is
\begin{eqnarray}
\beta F_{\rm p}~&\simeq&~N_{\rm +}\left[\ln(n_{\rm +}\Lambda^3)-1
\right]~+~N_{\rm -}\left[\ln(n_{\rm -}\Lambda^3)-1 \right] \nonumber \\ 
 &=&~N_{\mu}\left[\ln(n_{\mu}\Lambda^3)-1~+~x_{\rm +}\ln x_{\rm +}
~+~x_{\rm -}\ln x_{\rm -}\right],
\label{Fp3}
\end{eqnarray}
where $x_{\rm\pm}=N_{\rm\pm}/N_{\mu}$ are the mean microion concentrations.

\subsection{Linear Response Functions}

The linear response functions, $\chi_{ij}(k)$, $i,j=\pm$, 
of the two-component reference plasma are simply proportional to 
the corresponding partial structure factors, $S_{ij}(k)$:
\begin{equation}
\chi_{ij}(k)~=~-\beta n_{\mu}S_{ij}(k).
\label{chi2}
\end{equation}
Liquid state theory~\cite{HM2} now relates the partial structure factors to 
Fourier transforms of the pair correlation functions, $\hat h_{ij}(k)$, via
\begin{equation}
S_{ij}(k)~=~x_{i}\delta_{ij}+x_{i}x_{j}n_{\mu}\hat h_{ij}(k).
\label{Sij}
\end{equation}
The pair correlation functions are in turn related to Fourier transforms of 
the direct correlation functions, $\hat c_{ij}(k)$, by the 
Ornstein-Zernike (OZ) equation for mixtures:
\begin{equation}
\hat h_{ij}(k)~=~\hat c_{ij}(k)+ n_{\mu}\sum_l 
x_{l}\hat c_{il}(k)\hat h_{lj}(k), \qquad i,j,l=\pm
\label{hij}
\end{equation}
which serves in fact to define $\hat c_{ij}(k)$.
For such weakly-coupled plasmas as we encounter in charged colloids, 
the mean spherical approximation (MSA) provides a reasonable closure
for the OZ equation.
This amounts to approximating $c_{ij}(r)$ by its asymptotic 
($r \to \infty$) limit,  
$c_{ij}(r)\simeq -\beta v_{ij}(r)$, for all $r$, or equivalently, 
\begin{equation}
\hat c_{ij}(k)~\simeq~-\beta\hat v_{ij}(k)~=~
-\frac{4\pi\beta z_{i}z_{j}e^2}{\epsilon k^2},
\label{MSA}
\end{equation}
where $z_{i},z_{j}=\pm z$.
Since, in the MSA, $\hat c_{\rm ++}(k)=\hat c_{\rm --}(k)=-\hat c_{\rm +-}(k)
\equiv \hat c(k)$, it follows directly from Eq.~(\ref{hij}) that
\begin{equation}
\hat h_{\rm ++}(k)~=~\hat h_{\rm --}(k)~=~-\hat h_{\rm +-}(k)~=~
\frac{\hat c(k)}{1-n_{\mu}\hat c(k)}.
\label{h++}
\end{equation}
Substituting Eqs.~(\ref{Sij}), (\ref{MSA}), and (\ref{h++}) 
into Eq.~(\ref{chi2}) yields $\chi_{ij}(k)$, from which we obtain
\begin{equation}
\chi_{\rm ++}(k)-\chi_{\rm +-}(k)~=~
-\frac{\beta n_{\rm +}}{1-n_{\mu}\hat c(k)}~=~
-\frac{\beta n_{\rm +}}{1+\kappa^2/k^2},
\label{chi3}
\end{equation}
\begin{equation}
\chi_{\rm +-}(k)-\chi_{\rm --}(k)~=~
\frac{\beta n_{\rm -}}{1-n_{\mu}\hat c(k)}~=~
\frac{\beta n_{\rm -}}{1+\kappa^2/k^2},
\label{chi4}
\end{equation}
and
\begin{equation}
\chi(k)~=~-\frac{\beta n_{\mu}}{1-n_{\mu}\hat c(k)}~=~
-\frac{\beta n_{\mu}}{1+\kappa^2/k^2},
\label{chi5}
\end{equation}
where 
\begin{equation}
\kappa\equiv\left(\frac{4\pi n_{\mu}z^2e^2}{\epsilon k_{\rm B}T}\right)^{1/2}
=\left(\frac{4\pi n^{(0)}_{\mu}z^2e^2}{(1-\eta)\epsilon k_{\rm B}T}
\right)^{1/2},
\label{kappa}
\end{equation}
and $n^{(0)}_{\mu}=N_{\mu}/V=n_{\mu}(1-\eta)$ is
the total nominal microion number density.
As will be seen below, the parameter $\kappa$ plays the role of
the Debye screening constant (inverse screening length) in 
the microion density profiles and in the effective pair interaction.  

\subsection{Microion Density Profiles}

Specifying the macroion-microion interaction amounts to determining 
the value of the parameter $\alpha$ in Eq.~(\ref{vm+-r}) 
that ensures vanishing microion densities inside the macroion cores.  
This in turn requires a calculation of the real-space microion density profiles.
The first step of this calculation is to Fourier transform Eq.~(\ref{vm+-r}), 
with the result
\begin{equation}
\hat v_{\rm m\pm}(k)~=~\mp\frac{4\pi Zze^2}{\epsilon k^2}\left[(1-\alpha)
\cos(k\sigma/2)+\alpha\frac{\sin(k\sigma/2)}{k\sigma/2}\right].
\label{vm+-k1}
\end{equation}
Now substituting Eqs.~(\ref{chi3}), (\ref{chi4}), and (\ref{vm+-k1}) into 
Eqs.~(\ref{rho+}) and Eqs.~(\ref{rho-}) gives for the $k\neq 0$ 
Fourier components of the microion densities 
\begin{equation}
\hat \rho_{\rm\pm}({\bf k})~=~\pm~x_{\rm\pm}\frac{Z}{z}\left(\frac{\kappa^2}
{k^2+\kappa^2}\right)\left[(1-\alpha)\cos(k\sigma/2)+\alpha
\frac{\sin(k\sigma/2)}{k\sigma/2}\right] 
~\sum_{j=1}^{N_{\rm m}}\exp(i{\bf k}\cdot{\bf R}_j),\quad k\neq 0,
\label{rho+-k}
\end{equation}
where the sum is over the positions ${\bf R}_j$ of the macroions.
Next inverse transforming Eq.~(\ref{rho+-k}), while respecting 
the $k \to 0$ limits, $\hat\rho_{\rm\pm}(0)=N_{\rm\pm}$, we obtain
\begin{equation}
\rho_{\rm\pm}({\bf r})~=~\left\{ \begin{array}
{l@{\quad\quad}l}
\rho_{\rm\infty}~\pm~x_{\rm\pm}{\displaystyle 
\sum_{j=1}^{N_{\rm m}}} \rho_{\rm >}(|{\bf r}-{\bf R}_j|), 
\qquad |{\bf r}-{\bf R}_j|>\sigma/2 \\
x_{\rm\pm}{\displaystyle \sum_{j=1}^{N_{\rm m}}}
\rho_{\rm <}(|{\bf r}-{\bf R}_j|), 
\qquad |{\bf r}-{\bf R}_i|<\sigma/2,
\end{array} \right.
\label{rho+-r1}
\end{equation}
where $\rho_{\rm\infty}=x_{\rm -}n_{\rm +}+x_{\rm +}n_{\rm -}$ is 
the bulk density of positive or negative microions (far from any macroion).  
Note that in general 
$\rho_{\rm\infty} \neq n_{\rm s}$, although $\rho_{\rm\infty} \to n_{\rm s}$ 
in the limit $x_{\rm\pm} \to 1/2$ ($n_{\rm c}/n_{\rm s} \to 0$). 
In Eq.~(\ref{rho+-r1}), $\rho_{\rm >}(r)$ and $\rho_{\rm <}(r)$ are 
single-macroion orbitals, given by
\begin{equation}
\rho_{\rm >}(r)~=~\frac{Z}{z}\frac{\kappa^2}{4\pi}\left[(1-\alpha)
\cosh(\kappa\sigma/2)+\alpha\frac{\sinh(\kappa\sigma/2)}{\kappa\sigma/2}\right]
~\frac{\exp(-\kappa r)}{r}, \qquad r>\sigma/2, 
\label{rho>r1}
\end{equation}
and
\begin{equation}
\rho_{\rm <}(r)~=~\frac{Z}{z}\frac{\kappa^2}{4\pi}
\left(-1+\alpha+\frac{\alpha}{\kappa\sigma/2}\right)
{\exp(-\kappa\sigma/2)}~\frac{\sinh(\kappa r)}{r}, \qquad r<\sigma/2.
\label{rho<r}
\end{equation}
Vanishing of $\rho_{\rm <}(r)$ for $r<\sigma/2$ is evidently ensured by setting
\begin{equation}
\alpha~=~\frac{\kappa\sigma/2}{1+\kappa\sigma/2}.
\label{alpha}
\end{equation}
Finally, substituting this choice for $\alpha$ back into Eq.~(\ref{rho>r1})
specifies the $r>\sigma/2$ orbital as
\begin{equation}
\rho_{\rm >}(r)~=~\frac{Z}{z}\frac{\kappa^2}{4\pi}
~\frac{\exp(\kappa\sigma/2)}{1+\kappa\sigma/2}
~\frac{\exp(-\kappa r)}{r}, \qquad r>\sigma/2, 
\label{rho>r2}
\end{equation}
which is automatically normalized to the correct number of counterions 
per macroion ($Z/z$).  The corresponding microion density profiles are
the linear combinations
\begin{equation}
\rho_{\rm\pm}({\bf r})~=~\rho_{\rm \infty}~\pm~x_{\rm\pm}\frac{Z}{z}
\frac{\kappa^2}{4\pi}~\frac{\exp(\kappa\sigma/2)}{1+\kappa\sigma/2}
~\sum_{j=1}^{N_{\rm m}}\frac{\exp(-\kappa|{\bf r}-{\bf R}_j|)}
{|{\bf r}-{\bf R}_j|}, \qquad |{\bf r}-{\bf R}_j|>\sigma/2.
\label{rho+-r2}
\end{equation}
Expression~(\ref{rho>r2}) is seen to be of precisely the same form as 
the Debye-H\"uckel expression for the density of electrolyte ions 
around a macroion~\cite{Hunter}.  A significant distinction lies, however, 
in the definition of the screening constant, $\kappa$.  
Whereas the Debye-H\"uckel $\kappa$ depends on the {\it nominal} bulk density 
of electrolyte ions, our $\kappa$ [Eq.~(\ref{kappa})] depends rather 
on the {\it effective} mean microion density $n_{\mu}$ 
(in the volume unoccupied by macroions).
The importance of redefining the usual $\kappa$ in this way, 
particularly for concentrated suspensions, has been noted previously 
by Russel and coworkers~\cite{Russel}.

\subsection{Effective Pair Interaction and Volume Energy}

We are now in a position to derive the main results of the paper.
Considering first the effective electrostatic pair interaction between 
pseudo-macroions, we proceed by substituting Eq.~(\ref{alpha}) 
into Eq.~(\ref{vm+-k1}), obtaining for the macroion-microion interaction
\begin{equation}
\hat v_{\rm m\pm}(k)~=~\mp\frac{4\pi Zze^2}{\epsilon k^2}\left(\frac{1}
{1+\kappa\sigma/2}\right)\left[\cos(k\sigma/2)+\kappa\frac{\sin(k\sigma/2)}{k}
\right].
\label{vm+-k2}
\end{equation}
Next substituting Eqs.~(\ref{chi5}) and (\ref{vm+-k2}) into Eq.~(\ref{vindk1})
yields the induced potential: 
\begin{equation}
\hat v_{\rm ind}(k)~=~-\frac{2\pi Z^2e^2}{\epsilon k^2}\left(\frac{1}
{1+\kappa\sigma/2}\right)^2 \left(\frac{\kappa^2}{k^2+\kappa^2}\right)
\left[1+\cos(k\sigma) + 2\kappa\frac{\sin(k\sigma)}{k} + 
\kappa^2\frac{1-\cos(k\sigma)}{k^2}\right].
\label{vindk2}
\end{equation}
Fourier transformation of Eq.~(\ref{vindk2}) is a straightforward calculation, 
with the result 
\begin{equation}
v_{\rm ind}(r)~=~\left\{ \begin{array}
{l@{\quad\quad}l}
\frac{\displaystyle Z^2e^2}{\displaystyle \epsilon}\left(
\frac{\displaystyle \exp(\kappa\sigma/2)}
{\displaystyle 1+\kappa\sigma/2}\right)^2~
\frac{\displaystyle \exp(-\kappa r)}{\displaystyle r} - 
\frac{\displaystyle Z^2e^2}{\displaystyle \epsilon r}, & r>\sigma \\
-\frac{\displaystyle Z^2e^2}{\displaystyle 2\epsilon r}\left(
\frac{\displaystyle 1}{\displaystyle 1+\kappa\sigma/2}\right)^2
\left[(2+\kappa\sigma)\kappa r-\frac{1}{2} \kappa^2 r^2\right], & r<\sigma.
\end{array} \right.
\label{vindr}
\end{equation}
Finally, substituting Eq.~(\ref{vindr}) into Eq.~(\ref{veffr1}), we obtain 
an explicit expression for the effective electrostatic pair potential: 
\begin{equation}
v_{\rm eff}(r)~=~\frac{Z^2e^2}{\epsilon}\left(\frac{\exp(\kappa\sigma/2)}
{1+\kappa\sigma/2}\right)^2~\frac{\exp(-\kappa r)}{r}, \qquad r>\sigma. 
\label{veffr2}
\end{equation}
This result is seen to be identical in form to the electrostatic part of 
the DLVO effective pair potential~\cite{DLVO}, which is usually derived 
by linearizing the Poisson-Boltzmann equation.
The only distinction between our pair potential and the DLVO potential lies in 
the definition of the screening constant, ours [Eq.~(\ref{kappa})] 
being a factor $(1-\eta)^{-1/2}$ larger than the usual DLVO $\kappa$
to account for exclusion of microions from the macroion cores.

Now the volume energy may be explicitly determined from Eq.~(\ref{Eo1}).  
It follows immediately from Eq.~(\ref{vindr}) that
\begin{equation}
\lim_{r\to 0} v_{\rm ind}(r)~=~-\frac{Z^2e^2}{\epsilon}
~\frac{\kappa}{1+\kappa\sigma/2},
\label{vindr0}
\end{equation}
from Eq.~(\ref{vindk2}) that
\begin{equation}
\lim_{k\to 0} \hat v_{\rm ind}(k)~=~-\left(\frac{Z}{z}\right)^2
\hat v_{\rm ++}(0)~+~\frac{\pi Z^2e^2\sigma^2}{\epsilon}
~\frac{1+\kappa\sigma/6}{1+\kappa\sigma/2}~+~
\frac{4\pi Z^2e^2}{\epsilon\kappa^2},
\label{vindk0}
\end{equation}
and from Eq.~(\ref{vm+-k2}) that
\begin{equation}
\lim_{k\to 0} \hat v_{\rm m+}(k)~=~-\frac{Z}{z}\hat v_{\rm ++}(0)~+~ 
\frac{\pi Zze^2\sigma^2}{2\epsilon}~\frac{1+\kappa\sigma/6}{1+\kappa\sigma/2}.
\label{vm+k0}
\end{equation}
Substituting Eqs.~(\ref{vindr0}) -- (\ref{vm+k0}) into Eq.~(\ref{Eo1}), 
and using approximation (\ref{Fp3}), we obtain for the volume energy
\begin{equation}
\beta E_{\rm 0}~=~N_{\rm +}\ln(n_{\rm +}\Lambda^3)
+ N_{\rm -}\ln(n_{\rm -}\Lambda^3)
- N_{\rm m}\frac{Z^2e^2\beta}{2\epsilon}\frac{\kappa}{1+\kappa\sigma/2} 
- \frac{1}{2}\frac{(N_{\rm +}-N_{\rm -})^2}{N_{\rm +}+N_{\rm -}},
\label{Eo2}
\end{equation}
neglecting irrelevant constants.  Note that the infinities associated with
the $k\to 0$ limits formally cancel one another, as they must~\cite{Note1}.
The first term on the right side of Eq.~(\ref{Eo2}) is 
the ideal-gas plasma free energy, discussed in Sec.~\ref{plasma}.  
The second term, which accounts for the electrostatic energy of interaction 
between the macroions and their screening clouds of counterions, is equivalent 
to one half the interaction energy were all the counterions to be placed at 
a radial distance $\kappa^{-1}$ from the centers of their respective macroions.
The final term corresponds to the $k\to 0$ limit in Eq.~(\ref{Eo1}).
Our result for the volume energy is very similar to that derived by 
van Roij {\it et al}.~\cite{vRH,vRDH} from a density-functional expansion, 
differing only in the manner in which exclusion of microions from 
the macroion cores is incorporated.
While Eq.~(\ref{Eo2}) incorporates excluded volume effects through 
a dependence of the screening constant [Eq.~(\ref{kappa})] on the effective 
microion density, van Roij {\it et al}. incorporate them through 
an additional term in the volume energy [Eq.~(61) in Ref.~\cite{vRDH}].

In closing this section, we remark on the range of validity of the theory.  
First, although the linear response approximation presupposes relatively weak 
microion response to the macroions, and thus weak screening, the general form 
of the screened-Coulomb pair potential in bulk suspensions is broadly supported
by Poisson-Boltzmann cell model calculations~\cite{Alexander}, {\it ab initio} 
simulations~\cite{Lowen,abinitio}, and experiments~\cite{Grier}.
Second, the excluded volume corrections incorporated in the modified screening 
constant, $\kappa$, may become significant even in the weak-screening regime
for concentrated suspensions of weakly charged macroions. 
Finally, although the theory neglects (in mean-field fashion) fluctuations 
and correlations in the microion densities, Monte Carlo simulations 
and cell model calculations~\cite{Jonsson} for spherical macroions suggest 
that such correlations contribute only marginally to the total free energy.

\subsection{Osmotic Pressure and Bulk Modulus}

Being independent of the macroion coordinates, the microion volume energy,  
$E_{\rm 0}$, appears simply as an additive term in the total Helmholtz 
free energy of the system: $F=F_{\rm m}+E_{\rm 0}$, where $F_{\rm m}$ is 
the free energy of the equivalent one-component system of pseudo-macroions 
interacting via the effective pair potential $v_{\rm eff}(r)$.
Correspondingly, any thermodynamic quantity derived from this free energy
may be decomposed into effective macroion and microion contributions.
Since $E_{\rm 0}$ [Eq.~(\ref{Eo2})] depends on the mean macroion density 
-- both explicitly and implicitly through $\kappa$ -- 
it can significantly influence thermodynamic properties of the system, 
especially at low salt concentrations~\cite{vRDH,vRE,Graf,Denton1}.

As illustrations, we consider the osmotic pressure and bulk modulus.
A colloidal suspension in equilibrium, through a semi-permeable membrane, 
with a reservoir of salt solution exerts an osmotic pressure, 
$\Pi=P-P_{\rm r}$, defined as the difference between the pressure of 
the system, $P$, and that of the reservoir, $P_{\rm r}$.  
Treating the reservoir as an ideal gas of $N_{\rm r}$ salt ion pairs
in a volume $V_{\rm r}$, we have $\beta P_{\rm r}=2N_{\rm r}/V_{\rm r}$.
Chemical equilibrium is characterized by equality of the chemical potentials 
of salt ion species exchanged between the system and the reservoir.
The chemical potential of the salt, defined as the change in free energy 
upon adding a salt ion, includes a contribution arising from 
the effect of salt concentration on the macroion-macroion interaction
(through $\kappa$).  Thus, in general, the salt concentrations of the
system and reservoir are nontrivially related.
However, for systems sufficiently dilute that the macroion contribution
may be ignored -- an assumption we make here -- the condition for 
chemical equilibrium may be approximated by equality of 
the reservoir salt density, $N_{\rm r}/V_{\rm r}$, and the 
{\it effective} salt density of the system, $n_{\rm s}=(N_{\rm s}/V)/(1-\eta)$.
Note that the effective salt density exceeds the nominal salt density,  
$N_{\rm s}/V$, by the ratio of the total volume to the free volume 
unoccupied by the macroion cores.
The distinction here between nominal and reservoir salt densities is akin 
to that between nominal and reservoir polymer densities 
in colloid-polymer mixtures~\cite{Ilett}.
The reservoir pressure is then given by
\begin{equation}
\beta P_{\rm r}~\simeq~2n_{\rm s}.
\label{osmo1}
\end{equation}
The total pressure (or equation of state) of the system, 
$P=P_{\rm m}+P_{\rm 0}$, comprises a macroion contribution, $P_{\rm m}$, 
and a microion contribution, 
$P_{\rm 0}=-(\partial E_{\rm 0}/\partial V)_{N_{\rm m},N_{\rm s}}$.
Combining Eqs.~(\ref{Eo2}) and (\ref{osmo1}), we obtain
\begin{equation}
\beta\Pi\sigma^3~=~\beta P_{\rm m}\sigma^3~+~n_{\rm c}\sigma^3~-~
\frac{1}{16\pi}\frac{Z}{z^2}~\frac{(\kappa\sigma)^3}{(1+\kappa\sigma/2)^2}.
\label{osmo2}
\end{equation}
The same result is obtained for arbitrary macroion concentration in the limit 
of zero salt concentration ($n_{\rm s} \to 0$), in which case $P_{\rm r}=0$.
The second and third terms in Eq.~(\ref{osmo2}) represent, respectively, 
the ideal-gas pressure of the counterions and a van der Waals-like adjustment
that accounts for the attraction between counterions and macroions. 
With increasing counterion density, these two terms compete with each other, 
the attractive term acting to reduce the total osmotic pressure.
For weak microion screening ($\kappa\sigma<1$), where the electrostatic fields 
of the macroions are relatively weak and the microion densities 
close to uniform, Eq.~(\ref{osmo2}) should reasonably approximate 
the osmotic pressure.  In fact, in the limit $\kappa\sigma \to 0$, our result 
naturally tends to the correct ideal-gas limit, $\beta P_{\rm 0} \to n_{\rm c}$.
For stronger screening ($\kappa\sigma>1$), where the microion densities
are more nonuniform, nonlinear response effects may become significant.

The bulk modulus (or inverse compressibility), defined by 
$B\equiv -V(\partial \Pi/\partial V)_{N_{\rm m},N_{\rm s}}$, 
may be similarly expressed in the form $B=B_{\rm m}+B_{\rm 0}$, 
where $B_{\rm m}$ and $B_{\rm 0}$ are the macroion and microion contributions, 
respectively.  From Eq.~(\ref{osmo2}), we immediately obtain
\begin{equation}
\beta B\sigma^3~=~\beta B_{\rm m}\sigma^3~+~\frac{n_{\rm c}\sigma^3}{1-\eta}
~-~\frac{3}{32\pi}\frac{Z}{z^2}\frac{1}{1-\eta}
~\frac{(\kappa\sigma)^3(1+\kappa\sigma/6)}{(1+\kappa\sigma/2)^3},
\label{bulkm}
\end{equation}
which again includes repulsive and attractive microion terms.
In Eqs.~(\ref{osmo2}) and (\ref{bulkm}), the macroion contributions, 
$P_{\rm m}$ and $B_{\rm m}$, are understood to be obtained from 
a theory (or simulation) of a one-component system of particles 
interacting via the effective pair potential 
[Eqs.~(\ref{kappa}) and (\ref{veffr2})].  In practice, the macroion charge, 
notoriously difficult to extract from experiment, is usually replaced by
an adjustable parameter, the effective or renormalized charge, 
$Z^*$~\cite{Alexander}.

As a test of our results, we compare, in Fig.~1, the osmotic pressure 
predicted from Eq.~(\ref{osmo2}) with the recent experimental measurements 
of Reus {\it et al}.~\cite{Reus} for a colloidal fcc crystal in a 
highly deionized ($n_{\rm s}=0$) aqueous solvent at room temperature 
(Bjerrum length $\lambda_{\rm B}\equiv\beta z^2e^2/\epsilon=0.714$ nm).
As an approximation for the macroion pressure, $P_{\rm m}$, we use results 
of integral-equation calculations based on the virial equation with HNC 
closure for the liquid-state pair distribution function~\cite{Reus,Belloni}.
For the effective macroion charge, we take the value $Z^*=700$ 
estimated by Reus {\it et al}. to best match their phase diagram to 
the simulations of Robbins {\it et al.}~\cite{Robbins}.  

The microion contribution is seen to make the dominant contribution to 
the total osmotic pressure and to substantially improve the agreement 
between the one-component model and experiment, particularly at lower
volume fractions ($\eta<0.07$).  The last term in Eq.~(\ref{osmo2}) clearly 
is essential to reduce the rapidly increasing counterion ideal-gas pressure. 
The discrepancies at higher volume fractions ($\eta>0.07$) might be 
attributed, at least partially, to an underestimate of $P_{\rm m}$
by the liquid-state theory.  They may also reflect nonlinear response 
of the counterions and associated effective many-body interactions between
pseudo-macroions.  Future work will address influences of effective
triplet interactions on the osmotic pressure~\cite{Denton2}. 
It should be mentioned that the Poisson-Boltzmann cell model~\cite{Levin,Reus} 
(upper curve in Fig.~1) matches the experimental data well, especially at
higher $\eta$.  However, while cell models, which consider the distribution 
of microions within a Wigner-Seitz cell centered on a single macroion, 
are limited to periodic crystals, the more general one-component model 
applies to any thermodynamic phase.

A more stringent test of the theory is presented by the bulk modulus --
the curvature, with respect to density, of the free energy density.  
In recent experiments, Weiss {\it et al}.~\cite{Weiss} determined the bulk 
modulus of colloidal fcc crystals suspended in a deionized, aqueous solvent 
at room temperature ($\lambda_{\rm B}=0.714$ nm) by measuring the 
long-wavelength limit of the static structure factor.
For two samples, distinguished by nearest-neighbor distances,  
$a=(3/4\pi n_{\rm m})^{1/3}=2.5$ $\mu$m and $3.25$ $\mu$m, the measured
bulk moduli were argued to be lower than the predictions of DLVO theory, 
as estimated on the basis of an approximate elastic theory for 
the macroion contribution, $B_{\rm m}$.
For the denser crystal, the measured value was $B=0.016 \pm 0.005$ Pa,  
less than a third of the estimated ``DLVO" value of $B=0.052 \pm 0.005$ Pa.
This analysis ignores, however, the counterion contribution 
associated with the volume energy. 
Figures~2 and 3 present predictions, computed from Eq.~(\ref{bulkm}), 
for the counterion contribution, $B_{\rm 0}$. 
These results demonstrate that for sufficiently high effective 
macroion charge and volume fraction the counterion contribution may become 
{\it negative}.  It is essential to include this contribution 
in the total bulk modulus before comparing the DLVO theory with experiment.
In Fig.~2, the cross-over point at $Z^*\simeq 7100$ may be compared with
the effective charges, $Z^*\simeq 6100$ for $a=2.5$ $\mu$m and $Z^*\simeq 5200$ 
for $a=3.25$ $\mu$m, estimated by Weiss {\it et al.} for isolated pairs of
spheres in the infinite dilution limit.  However, lacking reliable knowledge 
of $Z^*$ in the crystal phase, we forgo here a more quantitative analysis. 
The qualitative message is nevertheless clear: at sufficient concentration, 
the counterions may act to {\it lower} the bulk modulus, softening
or even destabilizing ($B<0$) the crystal.

\section{Conclusions}\label{conclusions}

In summary, by reducing a model colloidal suspension of charged hard-sphere
macroions and point microions to an equivalent one-component system, 
and approximating the microion response to the macroion charge using 
linear response (second-order perturbation) theory, we have derived 
an effective electrostatic pair interaction [Eq.~(\ref{veffr2})] and 
an associated microion volume energy [Eq.~(\ref{Eo2})].  The volume energy, 
which depends on the average macroion density, accounts for both  
the microion entropy and the macroion-microion interaction energy. 
The effective interaction, which governs the dynamics of the macroions, 
is of precisely the conventional DLVO form for finite-sized macroions, 
but incorporates excluded volume corrections through the dependence of 
the screening constant on the {\it effective} density of microions 
in the free volume between macroion cores.  

The total free energy of the system is the sum of the volume energy and 
the free energy of the equivalent one-component system of pseudo-macroions. 
From the free energy, we have derived simple analytic expressions for
the osmotic pressure and bulk modulus.
Comparison of theoretical predictions with experimental data for deionized 
suspensions of highly charged macroions shows that the microions can 
significantly contribute to the thermodynamic properties, beyond their role 
in screening the bare Coulomb interaction between macroions.
In particular, the volume energy largely accounts for the observed magnitude 
of the osmotic pressure and qualitatively explains measurements of bulk 
modulus lower than predicted by the conventional one-component DLVO theory.
Several recent studies have predicted similar influences of volume energies 
on the phase behavior of charged colloids~\cite{vRH,vRDH,vRE,Graf,DL}. 

The theory presented here can be straightforwardly generalized to include 
{\it nonlinear} response of microions~\cite{Denton2} and thereby used 
to assess the relative importance of effective {\it many-body} 
interactions~\cite{Lowen,abinitio,triplet} and associated corrections 
to the effective pair potential and the volume energy.
Related applications are to colloid-surface interactions and to interactions 
between colloids in the vicinity of a surface, which experiment~\cite{Grier}
and theory~\cite{Goulding} suggest may become attractive.
Work along these lines is in progress.

\acknowledgments
I am grateful to Anne M.~Denton, Hartmut L\"owen, Hartmut Graf, and 
Christos N.~Likos for helpful discussions, and to Luc Belloni
for kindly supplying the HNC and PBC data used in Figure 1.

\bigskip
\noindent
*~Permanent address: Dept.~of Physics, North Dakota State University, 
Fargo, ND 58105

\appendix





\unitlength1mm

\begin{figure}
\bigskip
\bigskip
\begin{center}
\begin{picture}(120,90)
\put(5,0){\psfig{figure=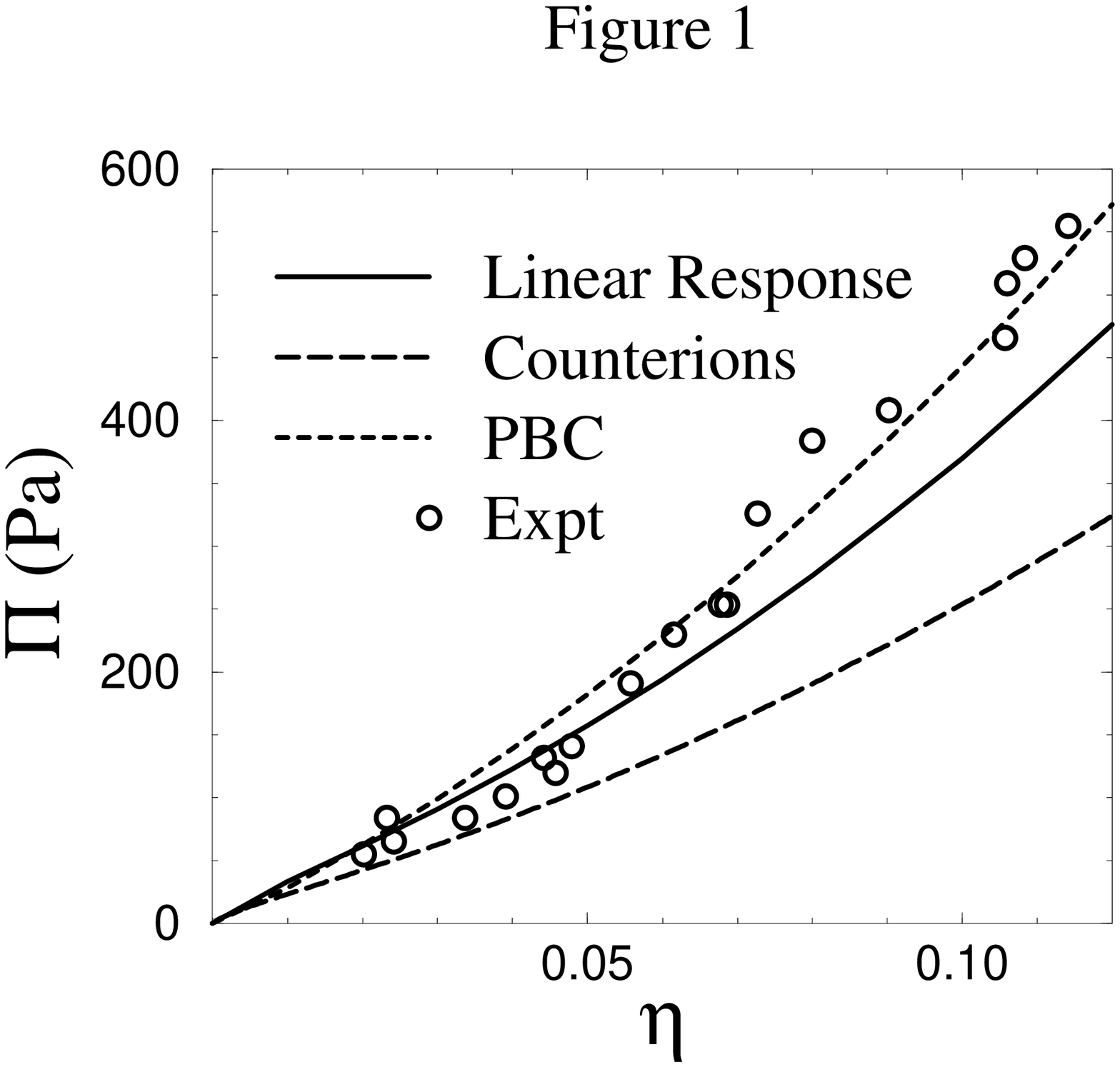,width=120mm,height=90mm}}
\end{picture}
\end{center}
\bigskip
\caption[]{
Osmotic pressure $\Pi$ vs. macroion volume fraction $\eta$ for an fcc crystal
of spherical macroions (diameter $\sigma=102$ nm) suspended in a salt-free,  
aqueous solvent at room temperature ($\lambda_{\rm B}=0.714$ nm).
Symbols: experimental data of Reus {\it et al}.~\cite{Reus}; 
solid curve: prediction of linear response theory [Eq.~(\ref{osmo2})]
with effective macroion charge $Z^*=700$ and HNC-virial pressure for
macroions (see text); long-dashed curve: counterion contribution $\Pi_{\rm 0}$;
short-dashed curve: Poisson-Boltzmann cell model 
prediction~\cite{Levin,Reus}.
Over this range of volume fractions, the screening constant increases 
from zero to $\kappa\sigma\simeq 4$.
}
\label{FIG1}
\end{figure}

\begin{figure}
\bigskip
\bigskip
\begin{center}
\begin{picture}(120,90)
\put(5,0){\psfig{figure=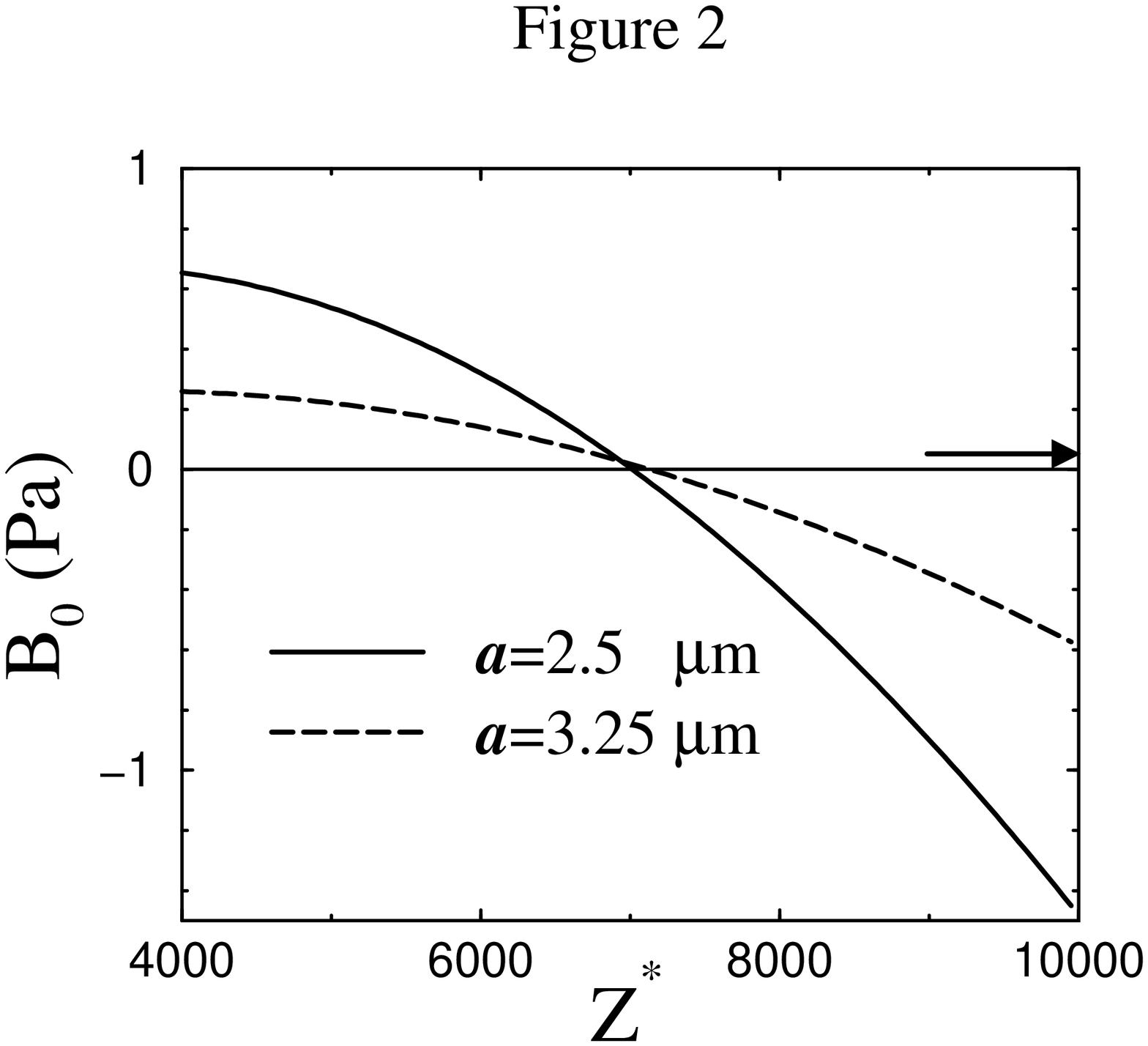,width=120mm,height=90mm}}
\end{picture}
\end{center}
\bigskip
\caption[]{
Counterion contribution to bulk modulus $B_{\rm 0}$ vs. effective macroion
charge $Z^*$ for fcc crystals of spherical macroions (diameter $\sigma=654$ nm)
suspended in a salt-free, aqueous solvent at room temperature.
Parameters are chosen for comparison with ref.~\cite{Weiss}.
Solid curve: nearest-neighbor distance $a=2.5$ $\mu$m 
(macroion volume fraction $\eta=0.0133$); 
dashed curve: $a=3.25$ $\mu$m ($\eta=0.00475$).
For $Z^*>7100$ the counterion contribution is negative. 
The arrow indicates the estimated macroion contribution to the bulk modulus
of the denser crystal~\cite{Weiss} (see text).
The maximum screening constant is $\kappa\sigma=1.9$. 
}
\label{FIG2}
\end{figure}

\begin{figure}
\bigskip
\bigskip
\begin{center}
\begin{picture}(120,90)
\put(5,0){\psfig{figure=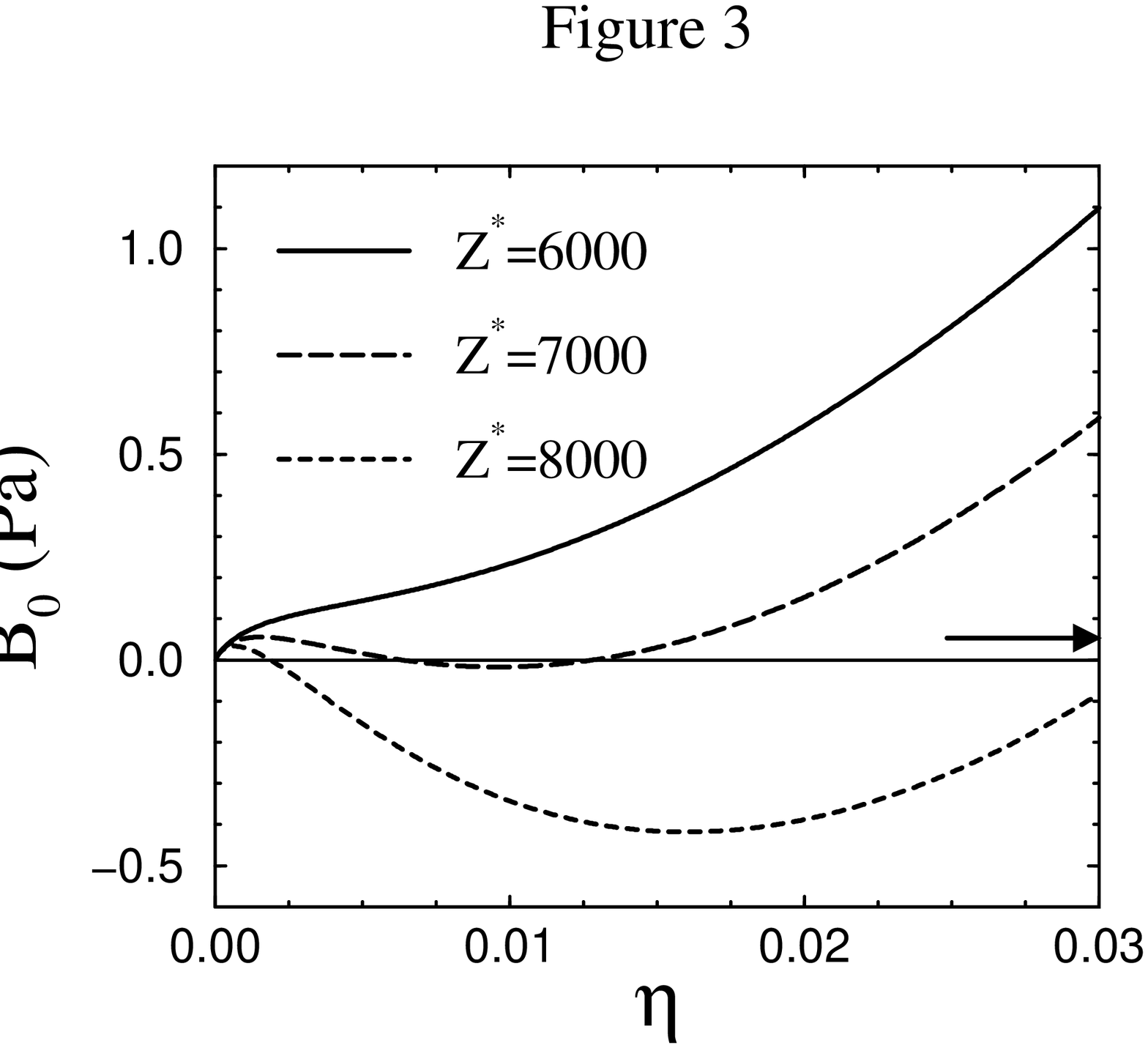,width=120mm,height=90mm}}
\end{picture}
\end{center}
\bigskip
\caption[]{
Counterion contribution to bulk modulus $B_{\rm 0}$ vs. macroion 
volume fraction $\eta$ for same system as in Fig.~2, but for three
different effective macroion charges.  Solid curve: $Z^*=6000$;  
long-dashed curve: $Z^*=7000$; short-dashed curve: $Z^*=8000$.
For sufficiently high $Z^*$, the counterion contribution may be negative
over a significant range of volume fractions.
The arrow indicates the estimated macroion contribution to the bulk modulus
of the denser crystal~\cite{Weiss}.
The maximum screening constant is $\kappa\sigma=2.5$.
}
\label{FIG3}
\end{figure}

\end{document}